\documentclass[12pt,preprint]{aastex}
\usepackage{graphics,graphicx,rotating,amsmath}

\newcommand{\lr}[1]{\left( #1 \right)}
\newcommand{\lrs}[1]{\left[ #1 \right]}
\newcommand{\lrt}[1]{\left< #1 \right>}
\newcommand{\bbe}{{\boldsymbol \beta}}
\newcommand{\bth}{{\boldsymbol \theta}}

\newcommand{\mbA}{{\mbox{\boldmath$A$}}}
\newcommand{\mbe}{{\mbox{\boldmath$e$}}}

\newcommand{\rSN}{{\rm SNR}}

\newcommand{\rSE}{{\rm SER}}
\newcommand{\tmbe}{{\tilde\mbe}}
\newcommand{\Dmbe}{{\Delta \mbe}}
\newcommand{\Dtmbe}{{\Delta \tilde\mbe}}

\newcommand{\mbetrue}{{\mbe_{g}}}

\newcommand{\ReS}{{S^{Re}}}
\newcommand{\ReN}{{N^{Re}}}

\newcommand{\tilG}{{\tilde G}}
\newcommand{\tilI}{{\tilde I}}
\newcommand{\tilP}{{\tilde P}}

\newcommand{\tilS}{{\tilde S}}
\newcommand{\tilN}{{\tilde N}}
\newcommand{\tilR}{{\tilde R}}
\newcommand{\tilReS}{{\tilde S^{Re}}}

\newcommand{\tilW}{{\tilde W}}

\newcommand{\cM}{{\cal M}}
\newcommand{\cS}{{\cal S}}

\newcommand{\MOM}[2]{{\cM^{#1}_{#2}}}

\newcommand{\SHP}[2]{{\cS^{#1}_{#2}}}

\newcommand{\CoefW}{{C_W}}
\newcommand{\CoefP}{{C_P}}
\newcommand{\eqG}{{\doteq}}
\newcommand{\RWRW}[1]{{Q\lr{#1}}}

\begin{document}
\title{Pixel Noise Effect and its Correction method by ERA Method for Precise Weak Gravitational Lensing Shear Measurement}
\author{Yuki Okura\altaffilmark{1}} 
\email{yuki.okura@riken.jp}

\author{Toshifumi Futamase\altaffilmark{2}}
\email{tof@tohoku.astr.ac.jp}

\altaffiltext{1}
 {RIKEN}
\altaffiltext{2}
 {Kyoto Sangyo University}

\begin{abstract}
Highly precise weak lensing shear measurement is required for statistical weak gravitational lensing analysis such as cosmic shear measurement to achieve severe constrain on the cosmological parameters. For this purpose any systematic error in the measurement should be corrected. One of the main systematic error comes from Pixel noise which is Poisson noise of flux from atmosphere.  
We investigate how the pixel noise makes systematic error in shear measurement based on ERA method and develop the correction method. 
This method is tested by simulations with various conditions and it is confirmed  
that the correction method can correct $80 \sim 90\%$ of the systematic error except very low signal to noise ratio galaxies. 
\end{abstract}

\section{Introduction}
It is widely recognized that weak gravitational lensing shear analysis is an unique and powerful tool to analyze the mass distribution of the universe.
Coherent deformation of the shapes of background galaxies carries not only the information of intervening mass distribution but also the cosmological background geometry and thus the cosmological parameters (Mellier 1999; Schneider 2006; Munshi et al. 2008).

In particular the cosmic shear which is the weak lensing by large scale structure of the universe has attracted much attention for the measurement of dark energy which is believed to be the reason of accelerated expansion of the universe. 
Although the cosmic shear have been measured by several groups in the past(Bacon et al. 2000, 2003; Maoli et al. 2001; Refregier et al. 2002; Hamana et al. 2003; Casertano et al. 2003; van Waerbeke et al. 2005; Massey et al. 2005; Hoekstra et al. 2006), the accuracy of the measurements are not enough to give a useful constraint on the dark energy 
parameter(the dark energy equation of state parameter $w$). 
For more precise and planned measurements, several wide survey observation were started such as 
Canada-France-Hawaii Telescope Legacy Survey(CFHTLS : http://www.cfht.hawaii.edu/Science/CFHTLS/), 
Dark Energy Survey(DES : https://www.darkenergysurvey.org/), 
Hyper Suprime-Cam(HSC : http://www.naoj.org/Projects/HSC/HSCProject.html) 
and 
The Large Synoptic Survey Telescope(LSST : http://www.lsst.org), 
EUCLID(EUCLID : http://sci.esa.int/euclid), 
Wide Field Infrared Survey Telescope(WFIRST : https://wfirst.gsfc.nasa.gov/).
Especially, the Hyper Suprime-Cam Subaru Strategic Program(SSP) plans 1400 deg$^2$ wide survey observation for constraining the cosmological parameters with lower than 1$\%$ error.
The observation has started in 2014, then recently about 100deg$^2$ of HSC wide survey data was published.
To achieve a severe constraint on the cosmic equation of state parameter dark energy, 
the HSC SSP requires high precise weak gravitational lensing shear analysis method with lower than $1\%$ systematic error.

Various methods of weak lensing shear analysis have been  developed for the precise weak gravitational lensing shear analysis such as Bayesian Fourier domain (Bernstein 2014),
Metacalibration method(Sheldon 2017, Huff 2017)m and some of them were tested with realistic simulation(Heymans et al 2006, Massey et al 2007, Bridle et al 2010 and Kitching et al 2012)
We have also developed a new weak lensing shear analysis method called E=HOLICs 
 (Okura and Futamase 2011, 2012, 2013) based on 
KSB(Kaiser et al. 1995) method, 
The method uses an elliptical weight function in order to 
avoid for the systematic error coming from the approximation of the weight function. 
We have then develop a new method of PSF correction called   
ERA (Okura and Fuatase 2014, 2015, 2016). 
Th method resmears galaxy and Point Spread Function(PSF) image to have idealized PSF which has ellipticity same as true ellipticity, Although ERA was able to eliminate some systematic error with enough precision, the systematic error by the pixel noise is not yet corrected. 
The pixel noise  is the Poisson noise of sky count, so it is random count on the observed galaxy image and thus changes the ellipticity of the galaxy image.
Even the noise count is random, the change of ellipticity has not only random component but also has systematic component which bring about the systematic error in measuring ellipticity. 

. 
Here we further develop ERA method to incorporate a  new method of PSF correction 
which take care of the systematic error caused by Pixel noise
in order to achieve required accuracy for cosmic shear measurement.  

This paper is organized as follows.
In section 2, we give a brief introduction of the basics of the ERA method.
In section 3, we derive how to correct pixel noise effect analytically based on the ERA method. 
In section 4, we test the correction method by simple simulation.
In section 5, we summarize and discuss our results.

\section{The Basics of The ERA Method}
The ERA method provides a new method of PSF correction by introducing 
Re-Smearing function(RSF). The RSF re-smears the observed galaxy and star images again to reshape PSF into the idealized PSF which has the same ellipticity with the lensed galaxy before the atmospheric smearing.   
The detailed explanation of ERA method can be seen in Okura and Futamase 2016.

\subsection{The notation and definitions}
ERA method make use of two planes. 
One is the lens plane, in this plane galaxy has ellipticity ``$\mbetrue$'' which is 
the distorted ellipticity distorted by gravitational lensing shear. We denote the position 
angle in this pale by ``$\bth''$ with the origin at the image centroid. 
Another is the zero plane where galaxy has 0 ellipticity, and we denote the position angle as ``$\bbe$'' in this plane with the origin at the image centroid.
The positions in two planes relate as   
\begin{eqnarray}
\bbe &=& \lr{1 - \kappa}\lr{\bth - \mbetrue\bth^*} = \mbA\bth\\
\bth &=& \frac{\bbe + \mbetrue\bbe^*}{\lr{1 - \kappa}\lr{1-|\mbetrue|^2}} = \mbA^{-1}\bbe,
\end{eqnarray}
An arbitrary ellipticity $\mbe_l$ in the lens plane and $\mbe_z$ in the zero plane is related as
\begin{eqnarray}
\label{eq:nonle}
\mbe_l &=& \frac{\mbe_z + \mbetrue}{1 + \mbe_z\mbetrue^*}
\end{eqnarray}
From the definition of the zero plane, each galaxies have different zero planes each other.

Let ``$G(\bth)$'' and ``$\tilG(\bbe)$'' be the brightness distribution of galaxy in the 
lens and the zero plane, respectively. 
The image moments of the galaxy in the zero plane is defined as
\begin{eqnarray}
\MOM{N}{M}(\tilG) = \int d^2\beta \bbe^N_M \tilG(\bbe) W\lr{{\bbe^2_0}/{\sigma_W^2}},
\end{eqnarray}
where
\begin{eqnarray}
\bbe^N_M&\equiv&\bbe^{\frac{N+M}{2}}\bbe^{*\frac{N-M}{2}},
\end{eqnarray}
and W is the weight function which must be concentric function, e.g. circular Gaussian function,  in the zero plane, and $\sigma_W^2$ is the weight scale.
This moment can be measured in the lens plane as 
\begin{eqnarray}
\MOM{N}{M}(G) = \int d^2\theta \lr{\mbA\bth}^N_M G(\bth) W\lr{{\lr{\mbA\bth}^2_0}/{\sigma_W^2}},
\end{eqnarray}
where we ignored the scalar coefficient from Jacobian, it is not important for this study.

From the definition of the zero plane, galaxy image has centroid at the origin of the coordinate and has no ellipticity, so the dipole and the ellipticity must be 0.   
Thus the complex moments satisfy the following identities. 
\begin{eqnarray}
\label{eq:Mzero_1}
\MOM{1}{1}(\tilG) &=& \int d^2\beta \bbe^1_1 \tilG(\bbe) W\lr{{\bbe^2_0}/{\sigma_W^2}}
 = \int d^2\theta \lr{\mbA\bth}^1_1 G(\bth) W\lr{{\lr{\mbA\bth}^2_0}/{\sigma_W^2}} = 0\\
\label{eq:Mzero_2}
\MOM{2}{2}(\tilG) &=& \int d^2\beta \bbe^2_2 \tilG(\bbe) W\lr{{\bbe^2_0}/{\sigma_W^2}}
 = \int d^2\theta \lr{\mbA\bth}^2_2 G(\bth) W\lr{{\lr{\mbA\bth}^2_0}/{\sigma_W^2}} = 0.
\end{eqnarray}
One can determine the zero plane by finding the centroid and the ellipticity which satisfy the these equations \ref{eq:Mzero_1} and \ref{eq:Mzero_2}.

When the galaxy has the pixel noise with scale $\sigma_{PN}$, the signal-to-noise ratio $\rSN$ is defined as 
\begin{eqnarray}
\rSN \equiv \frac{\MOM{0}{0}(\tilG)}{\sigma_{PN}\sqrt{\MOM{0}{0}(\tilW)}},
\end{eqnarray}
where
\begin{eqnarray}
\MOM{N}{M}(\tilW) = \int d^2\beta \bbe^N_M W^2\lr{{\bbe^2_0}/{\sigma_W^2}} = \int d^2\theta \lr{\mbA\bth}^N_M W^2\lr{{\lr{\mbA\bth}^2_0}/{\sigma_W^2}},
\end{eqnarray}
and the weight scale is determined to maximize the signal-to-noise ratio, so the weight scale is determined from 
${\partial\rSN}/{\partial\sigma_W^2} = 0$. The condition can be described as
\begin{eqnarray}
\frac{\MOM{'2}{0}(\tilG)}{\MOM{0}{0}(\tilG)}
-
\frac{\MOM{'2}{0}(\tilW)}{\MOM{0}{0}(\tilW)}
=0.
\end{eqnarray}
where $\MOM{'N}{M}(\tilG) = \int d^2\beta \bbe^N_M \tilG(\bbe) W'\lr{\bbe^2_0/\sigma_W^2}$ and $W'(x) = {\partial W(x)}/{\partial x}$. 
If the weight function is a Gaussian function, the weight scale $\sigma_W$ and Gaussian radius $r_g$ is determined as
\begin{eqnarray}
\label{eq:g_r}
\sigma_W^2 = 2r_g^2 = 2\frac{\MOM{2}{0}(\tilG)}{\MOM{0}{0}(\tilG)}.
\end{eqnarray}

\subsection{PSF correction}
ERA method  make use of the zero plane also for PSF correction. 
PSF effect smears galaxy image $G(\bth)$ by PSF ``$P(\bth)$'' and make a smeared galaxy image ``$S(\bth)$'' which is described in the zero plane as follows.
\begin{eqnarray}
S(\bth) = G(\bth)*P(\bth)
\end{eqnarray}
and 
\begin{eqnarray}
\tilS(\bbe) = \tilG(\bbe)*\tilP(\bbe)
\end{eqnarray}
where $*$ means convolution.

Let us imagine that the PSF has a concentric function in the zero plane. 
If so,  the smeared Galaxy also is concentric. 
This means that the smeared galaxy image satisfies equation \ref{eq:Mzero_2} instead of $G$ and has the ellipticity $\mbetrue$ in the lens plane, 
and so the true ellipticity $\mbetrue$ can be measured by simply measuring the shape of the smeared galaxy image.
However, in real analysis, PSF shape does not have such an idealized shape but rather  complicated shape and the shape changes in each exposures. 
ERA method creates such an ideal situation by re-smearing the smeared galaxy image again by re-smearing function. 

Let ``$I(\bbe^2_0/\sigma_I^2)$'' is the idealized PSF which is a concentric function in the zero plane and should have slightly large radius than PSF.
The Re-smeared function ``$\tilR(\bbe)$'' is then defined in the zero plane as
\begin{eqnarray}
\tilR(\bbe) = \tilI(\bbe^2_0/\sigma_I^2)\otimes\tilP(\bbe),
\end{eqnarray}
where $\otimes$ denotes the deconvolution.
Then the re-smeared galaxy ``$\tilReS(\bbe)$'' is defined as
\begin{eqnarray}
\tilReS(\bbe) = \tilS(\bbe)*\tilR(\bbe) = \tilG(\bbe)*\tilI(\bbe^2_0/\sigma_I^2),
\end{eqnarray}
and also $\tilReS(\bbe)$ satisfies equation \ref{eq:Mzero_2} as
\begin{eqnarray}
\label{eq:Mzero_2RES}
\MOM{2}{2}(\tilReS)
 &=& \int d^2\beta \bbe^2_2 \tilReS(\bbe) W\lr{{\bbe^2_0}/{\sigma_W^2}}
\nonumber\\
 &=& \int d^2\beta \bbe^2_2 \lr{\tilS(\bbe)*\tilR(\bbe)} W\lr{{\bbe^2_0}/{\sigma_W^2}}
\nonumber\\
 &=& \int d^2\beta \bbe^2_2 \lr{\tilG(\bbe)*I(\bbe^2_0/\sigma_I^2)} W\lr{{\bbe^2_0}/{\sigma_W^2}}
\nonumber\\
 &=& \int d^2\theta \lr{\mbA\bth}^2_2 \lr{G(\bth)*I((\mbA\bth)^2_0/\sigma_I^2)} W\lr{{\lr{\mbA\bth}^2_0}/{\sigma_W^2}}
\nonumber\\
 &=& \int d^2\theta \lr{\mbA\bth}^2_2 \lr{S(\bth)*R(\mbA\bth)} W\lr{{\lr{\mbA\bth}^2_0}/{\sigma_W^2}}
 = 0,
\end{eqnarray}
Therefore the ellipticity $\mbetrue$ which satisfies this equation is the PSF corrected ellipticity.

\section{Pixel Noise correction}
Pixel noise is Poisson noise of flux from atmosphere, this adds random count ``$N(\bth)$'' for image of objects.
In this paper, we assume star images for PSF measurement has high brightness enough to neglect pixel noise effect, 
so only observed(smeared by PSF and has pixel noise) galaxy ``$O(\bth)$'' is described as
\begin{eqnarray}
\label{eq:OBS}
O(\bth) = S(\bth) + N(\bth).
\end{eqnarray}
From the property of random count, the followings are obtained
\begin{eqnarray}
\label{eq:PRND}
\lrt{N(\bth)}&=&0\\
\lrt{N(\bth)N(\bth')}&=&\sigma^2_{PN}\delta(\bth-\bth'),
\end{eqnarray}
where the bracket means taking average for enough number of different random count fields.
In zero plane, same equations are obtained as
\begin{eqnarray}
\label{eq:PRND_zero}
\lrt{\tilN(\bbe)}&=&0\\
\lrt{\tilN(\bbe)\tilN(\bbe')}&=&\sigma^2_{PN}\delta(\bbe-\bbe'),
\end{eqnarray}
because moving to other plane is just 2-dimensional spatial scale change.

\subsection{Pixel Noise effect for ellipticity measurement without PSF correction}
The pixel noise count changes the shape of galaxy and so changes the ellipticity of galaxy.
Let $\mbe'$ and $\tmbe'$ be the observed ellipticity from a galaxy image with pixel noise in the lens plane and in the zero plane, and 
$\Dmbe$ and $\Dtmbe$ be the additional changes of ellipticitiies from the true values , respectively. Thus we have  
\begin{eqnarray}
\tmbe' &=& \Dtmbe\\
\mbe'  &=& \mbetrue + \Dmbe
\end{eqnarray}
The relations between these ellipticities are derived as 
\begin{eqnarray}
\tmbe' &=& \frac{\mbe' - \mbetrue}{1 - \mbe'\mbetrue^*}\\
\mbe' &=& \frac{\tmbe' + \mbetrue}{1 + \tmbe'\mbetrue^*}\\
\label{eq:addte}
\Dtmbe &=& \frac{\Dmbe}{1 - |\mbetrue|^2 - \Dmbe\mbetrue^*}\equiv\frac{\Dmbe'}{1 - \Dmbe'\mbetrue^*}\\
\label{eq:adde}
\Dmbe' &\equiv& \frac{\Dmbe}{1 - |\mbetrue|^2} = \frac{\Dtmbe}{1 + \Dtmbe\mbetrue^*}
\end{eqnarray}

Since the pixel noise changes ellipticity of galaxy, the corresponding zero plane(we call it as the zero plane with noise) is not the original one 
but is determine as follows. 
\begin{eqnarray}
\label{eq:shape_lens}
&&\int d^2\beta' \lr{\bbe'}^2_2 \lr{\tilG(\bbe)+\tilN(\bbe)} W((\bbe')^2_0/\sigma_W^2) = 
\nonumber\\
&&\int d^2\theta \lr{\mbA'\bth}^2_2 \lr{G(\bth)+N(\bth)} W((\mbA'\bth)^2_0/\sigma_W^2) = 0
\end{eqnarray}
where $\bbe' = \mbA'\bth$ is the coordinates in the zero plane with noise.
The coordinate can be written in terms of the observed ellipticity as $\mbA'\bth = \lr{1 - \kappa}\lr{\bth - \mbe'\bth^*}$, then it is related with $\bbe$ as
\begin{eqnarray}
\bbe' = \mbA'\bth  = \frac{\bbe-\Dtmbe\bbe^*}{1+\Dtmbe\mbetrue^*} = \bbe - \frac{\Dtmbe}{1+\Dtmbe\mbetrue^*} \lr{\bbe^*+\mbetrue^*\bbe},
\end{eqnarray}
Thus the equation \ref{eq:shape_lens} can be written in the original zero plane a follows.
\begin{eqnarray}
\label{eq:shape_zero}
\int d^2\beta \lr{\frac{\bbe-\Dtmbe\bbe^*}{1+\Dtmbe\mbetrue^*}}^2_2 \lr{\tilG(\bbe)+\tilN(\bbe)} W\lr{\lr{\frac{\bbe-\Dtmbe\bbe^*}{1+\Dtmbe\mbetrue^*}}^2_0/\sigma_W^2} = 0
\end{eqnarray}
The weight function is expanded up to the 2nd order in $\Dtmbe$ as follows
\begin{eqnarray}
\label{eq:expW}
W\lr{\lr{\frac{\bbe-\Dtmbe\bbe^*}{1+\Dtmbe\mbetrue^*}}^2_0/\sigma_W^2} 
&\approx&
W(\bbe^2_0/\sigma_W^2) 
\nonumber\\&&\hspace{-150pt}
-\frac{W'(\bbe^2_0/\sigma_W^2)}{\sigma^2_W}\lr{\lr{\lr{2\Dtmbe\cdot\mbetrue - \lr{1-|\mbetrue|^2}|\Dtmbe|^2-4\lr{\Dtmbe\cdot\mbetrue}^2}}\bbe^2_0
+2\lr{1-2\Dtmbe\cdot\mbetrue}\Dtmbe\cdot\bbe^2_2}
\nonumber\\&&
+\frac{2W''(\bbe^2_0/\sigma_W^2) }{\sigma_W^4}\lr{\Dtmbe\cdot\mbetrue\bbe^2_0+\Dtmbe\cdot\bbe^2_2}^2
\nonumber\\&&\hspace{-150pt}\approx
W(\bbe^2_0/\sigma_W^2) 
-\frac{2W'(\bbe^2_0/\sigma_W^2)}{\sigma^2_W}\lr{\lr{\Dtmbe\cdot\mbetrue}\lr{\bbe^2_0-\Dtmbe\bbe^2_{-2}}
+\Dtmbe\cdot\bbe^2_2}
+\frac{2\Dtmbe W''(\bbe^2_0/\sigma_W^2) }{\sigma_W^4}\lr{\Dtmbe\cdot\mbetrue}\bbe^4_{-2}
\end{eqnarray}
where in the last approximation we neglect terms which make only 0 value moments in 
the further calculations and 
$2\Dtmbe\cdot\mbe = \Dtmbe^*\mbe + \Dtmbe\mbe^*$.
Thus the equation \ref{eq:shape_zero} can be approximated up to the second order in $\Dtmbe$ as
\begin{eqnarray}
\lrs{\Dtmbe\lr{-2-\frac{\SHP{'4}{0}}{\sigma_W^2}}+|\Dtmbe_{(1)}|^2\frac{\mbetrue}{\sigma_W^2}\lr{3\SHP{'4}{0}+\frac{\SHP{''6}{0}}{\sigma_W^2}}}_{(G)} + \lrs{\SHP{2}{2}-\frac{\mbetrue\Dtmbe^*\SHP{'4}{2}}{\sigma_W^2}}_{(N)} = 0,
\end{eqnarray}
where the moments are normalized by the quadrupole moment of $\tilG$,
so $\SHP{N}{M}(A) \equiv \MOM{N}{M}(A)/\MOM{2}{0}(G)$, 
By comparing 1st order and 2nd order, we obtain the following: 
\begin{eqnarray}
\label{eq:shape_zero1st}
\Dtmbe_{(1)} &=& \frac{1}{2+\frac{\SHP{'4}{0}(G)}{\sigma_W^2}}\SHP{2}{2}(N) \equiv \CoefW\SHP{2}{2}(N)  \\
\label{eq:shape_zero2nd}
\Dtmbe_{(2)} &=& \CoefW\mbetrue\lr{|\Dtmbe_{(1)}|^2\lr{\frac{3\SHP{'4}{0}(G)}{\sigma_W^2}+\frac{\SHP{''6}{0}(G)}{\sigma_W^4}}-\Dtmbe_{(1)}^*\frac{\SHP{'4}{2}(N)}{\sigma_W^2}}.
\end{eqnarray}
The statistical values of the 1st order and 2nd order additional ellipticity are thus  
\begin{eqnarray}
\lrt{\Dtmbe_{(1)}} &=& 0\\
\lrt{|\Dtmbe_{(1)}|^2} &=& \CoefW^2\lrt{\SHP{2}{2}(N)\SHP{2}{-2}(N)} =
\lr{\CoefW\frac{\SHP{0}{0}(G)}{\rSN}}^2\frac{\MOM{4}{0}(W)}{\MOM{0}{0}(W)}\\
\lrt{\Dtmbe_{(1)}^2} &=& 0\\
\lrt{\Dtmbe_{(2)}} &=& 
\mbetrue\lr{\CoefW\frac{\SHP{0}{0}(G)}{\rSN}}^2
\lr{\CoefW\frac{\MOM{4}{0}(W)}{\MOM{0}{0}(W)}\lr{\frac{3\SHP{'4}{0}(G)}{\sigma_W^2}+\frac{\SHP{''6}{0}(G)}{\sigma_W^4}}-
\frac{1}{\sigma_W^2}\frac{\MOM{'6}{0}(W)}{\MOM{0}{0}(W)}} \eqG 0.
\end{eqnarray}
where $\lrt{}$ means the average value by taking enough number of different random count images,
\begin{eqnarray}
\lrt{\SHP{N}{M}(N)\SHP{O}{P}(N)}& = &
\lr{\frac{\sigma_{PN}^2}{\MOM{2}{0}(G)}}^2\lrt{\MOM{N}{M}(N)\MOM{O}{P}(N)}
\nonumber\\& = &
\lr{\frac{\SHP{0}{0}(G)}{\rSN}}^2\frac{\MOM{N+O}{M+P}(W)}{\MOM{0}{0}(W)},
\end{eqnarray}
where $\eqG$ means the analytical result in the case that the galaxy image and weight function are both elliptical Gaussian.
This result means that the pixel noise almost does not make systematic error and makes only concentric dispersion in zero plane, so it behaves like intrinsic ellipticity noise.

The additional ellipticity in the lens plane is obtained as 
\begin{eqnarray}
\lrt{\Dmbe} = \lr{1-|\mbe|^2}\lrt{\frac{\Dtmbe}{1+\Dtmbe\mbetrue}}
\approx
\lr{1-|\mbe|^2}\lrt{\Dtmbe_{(2)}}
\equiv
\Dmbe_{cor}
\end{eqnarray}

Finally, we can obtain Pixel Noise corrected ellipticity $\mbe_{cor}$ as follows.
\begin{eqnarray}
\mbe_{cor} = \mbe' - \Dmbe_{cor}
\end{eqnarray}

So the pixel noise effect for shape measurement can be corrected by simply taking non-linear average, i.e. the equation \ref{eq:nonle}, for the observed ellipticity and it is not needed any other correction terms.

We can define the weight of galaxies when measuring shear by averaging as
\begin{eqnarray}
w = \frac{\sigma_{int}^2}{\sigma_{int}^2 + \lrt{|\Dtmbe_{(1)}|^2}}
\end{eqnarray}
where $\sigma_{int}$ is the standard deviation of the intrinsic ellipticity.

\subsection{Pixel Noise effect for ellipticity measurement with PSF correction}
ERA method re-smears the galaxy image again by re-smearing function to reshape PSF to idealized PSF.
The idealized PSF has the same ellipticity as PSF corrected ellipticity which is affected by pixel noise, so the equation \ref{eq:Mzero_2RES} changes by pixel noise as follows
\begin{eqnarray}
\label{eq:Mzero_2RESN}
\MOM{2}{2}(\tilReS)
 &=& \int d^2\theta \lr{\mbA'\bth}^2_2 \lr{S(\bth) + N(\bth)}*R(\mbA'\bth) W\lr{{\lr{\mbA'\bth}^2_0}/{\sigma_W^2}}
 = 0,
\end{eqnarray}
where $R(\mbA'\bth) = I(\lr{\mbA'\bth}^2_0/\sigma_W^2)\otimes P(\bth)$ is the re-smearing function. 

The re-smearing function may be expanded up to the second order in  
  $\Dtmbe$ as in the weight function equation \ref{eq:expW}. 
Then  \ref{eq:Mzero_2RESN} gives us the following.
\begin{eqnarray}
\label{eq:shear_zero1st}
\Dtmbe_{(1)} &=& \frac{\SHP{2}{2}(\ReN)}{2+\frac{\SHP{'4}{0}(\ReS)}{\sigma_W^2}+\frac{\SHP{2}{2}(\ReS^{'2}_{-2})}{\sigma_R^2}} \equiv \CoefP\SHP{2}{2}(\ReN)  \\
\label{eq:shear_zero2nd}
\Dtmbe_{(2)} &=& 
\CoefP\Dtmbe_{(1)}\lr{2\Dtmbe_{(1)}\cdot\mbetrue}
\Biggl(
\frac{3\SHP{'4}{0}(\ReS)}{\sigma_W^2}
+\frac{\SHP{''6}{0}(\ReS)}{\sigma_W^4}
+\frac{2\SHP{2}{0}(\ReS^{'2}_{0})
+\SHP{2}{2}(\ReS^{'2}_{-2})}{\sigma_R^2}
\nonumber\\&&\hspace{150pt}
+\frac{\SHP{2}{2}(\ReS^{''4}_{-2})}{\sigma_R^4}
+\frac{\SHP{4}{0}(\ReS^{2}_{0})+\SHP{4}{2}(\ReS^{2}_{-2})}{\sigma_R^2\sigma_W^2}
\Biggr)
\nonumber\\&&\hspace{-50pt}-
\CoefP\Biggl(
\Dtmbe_{(1)}\lr{2\SHP{2}{0}(\ReN)
+\frac{\SHP{'4}{0}(\ReN)+\mbetrue^*\SHP{'4}{2}(\ReN)}{\sigma_W^2}
+\frac{\SHP{2}{2}(\ReN^{'2}_{-2})+\mbetrue^*\SHP{2}{2}(\ReN^{'2}_{0})}{\sigma_R^2}
}
\nonumber\\&&\hspace{0pt}
+\Dtmbe^*_{(1)}\lr{
\frac{\SHP{'4}{4}(\ReN)+\mbetrue\SHP{'4}{2}(\ReN)}{\sigma_W^2}
+\frac{\SHP{2}{2}(\ReN^{'2}_{2})+\mbetrue\SHP{2}{2}(\ReN^{'2}_{0})}{\sigma_R^2}
}\Biggr).
\end{eqnarray}
where, ${{\tilde S}^{ReN}_{M}} \equiv \tilS*(\tilR\beta^{N}_{M})$ and similar for ${{\tilde N}^{ReN}_{M}}$,
and shape $\SHP{N}{M}$ is normalized by quadrupole moments of $\tilReS$.
$\Dtmbe_{(1)}$ is the statistical noise and $\lrt{\Dtmbe_{(2)}}$ is the systematic noise from pixel noise in PSF corrected ellipticity
The dispersion and systematic error from the pixel noise can be predicted by taking average of equation \ref{eq:shear_zero1st} and \ref{eq:shear_zero2nd}. 
\begin{eqnarray}
\lrt{|\Dmbe|^2} &\approx& \lrt{|\Dmbe_{(1)}|^2}  = \lr{1-|\mbetrue|^2}^2\lrt{|\Dtmbe_{(1)}|^2}
\nonumber\\&=&
\CoefP^2\sigma^2_{PN}\lr{1-|\mbetrue|^2}^2{\RWRW{\tilR*W^2_2, \tilR*W^2_{-2}}}
\\
\lrt{\Dmbe} &\approx& \lrt{\Dmbe_{(2)}}
 = \lr{1-|\mbetrue|^2}\lrt{\Dtmbe_{(2)}}
\nonumber\\&=&
\CoefP^2\sigma^2_{PN}\lr{1-|\mbetrue|^2}
\Biggl(\CoefP\RWRW{\tilR*W^2_2, \tilR*W^2_{-2}\mbetrue + \tilR*W^2_2\mbetrue^*}
\times\nonumber\\&&\hspace{-50pt}
\Biggl(
\frac{3\SHP{'4}{0}(\ReS)}{\sigma_W^2}
+\frac{\SHP{''6}{0}(\ReS)}{\sigma_W^4}
+\frac{2\SHP{2}{0}(\ReS^{'2}_{0})
+\SHP{2}{2}(\ReS^{'2}_{-2})}{\sigma_R^2}
+\frac{\SHP{2}{2}(\ReS^{''4}_{-2})}{\sigma_R^4}
+\frac{\SHP{4}{0}(\ReS^{2}_{0})+\SHP{4}{2}(\ReS^{2}_{-2})}{\sigma_R^2\sigma_W^2}
\Biggr)
\nonumber\\&&\hspace{-75pt}-
\RWRW{\tilR*W^2_2, 2\tilR*W^2_0
+\frac{\tilR*W^{`4}_0 + \tilR*W^{`4}_2\mbetrue^*}{\sigma_W^2}
+\frac{\tilR^{`2}_{-2}*W^2_0 + \tilR^{`2}_0*W^2_2\mbetrue^*}{\sigma_R^2}
}
\nonumber\\&&\hspace{-75pt}-
\RWRW{\tilR*W^2_{-2}, \frac{
\tilR*W^{`4}_4 + \tilR*W^{`4}_2\mbetrue}{\sigma_W^2}+
\frac{\tilR^{`2}_2*W^2_2 + \tilR^{`2}_0*W^2_2\mbetrue}{\sigma_R^2}}
\Biggr)
\end{eqnarray}
where
\begin{eqnarray}
\RWRW{\tilR^N_M*W^O_P, \tilR^Q_R*W^S_T}&\equiv&\frac{\int d^2\beta \lr{\tilR^N_M*W^O_P}\lr{\tilR^Q_R*W^S_T}}{\lr{\MOM{2}{0}\lr{\ReS}}^2}\\
\tilR^N_M &\equiv& \lr{I(\bbe^2_0/\sigma^2_I)\bbe^N_M}\otimes\tilP(\bbe)\\
W^N_M &\equiv&W\lr{\bbe^2_0/\sigma^2_W}\bbe^N_M
\end{eqnarray}
\section{Simulation test}

We test the correction method derived in the previous section by simulation using  some simple models of galaxy and PSF. 

We use Gaussian and Sersic profiles for galaxy whose radius determined by equation \ref{eq:g_r} is 2.0 pixels and the ellipticity is 0.2.
For PSF we use Gaussian PSF, Gaussian weight function and Gaussian idealized PSF. ,
The radius of PSF is selected form [1.5, 2.0, 2.5] pixels and ellipticity from [0.0, -0.1, 0.1i].

The steps of the simulation is as follows. 
First we create galaxy and PSF image, and make the smeared galaxy image by convolving the galaxy and the PSF image. Then we add random count image for the smeared galaxy image. Finally we measure ``PSF corrected ellipticity'' and ``PSF and Pixel Noise corrected ellipticity''.  
We measure the two ellipticities with 40,000 different random count images, then we calculate the mean of the ellipticities with two times 5$\sigma$ outlier clipping.

\begin{figure*}[tbp]
\centering
\resizebox{1.0\hsize}{!}{\includegraphics{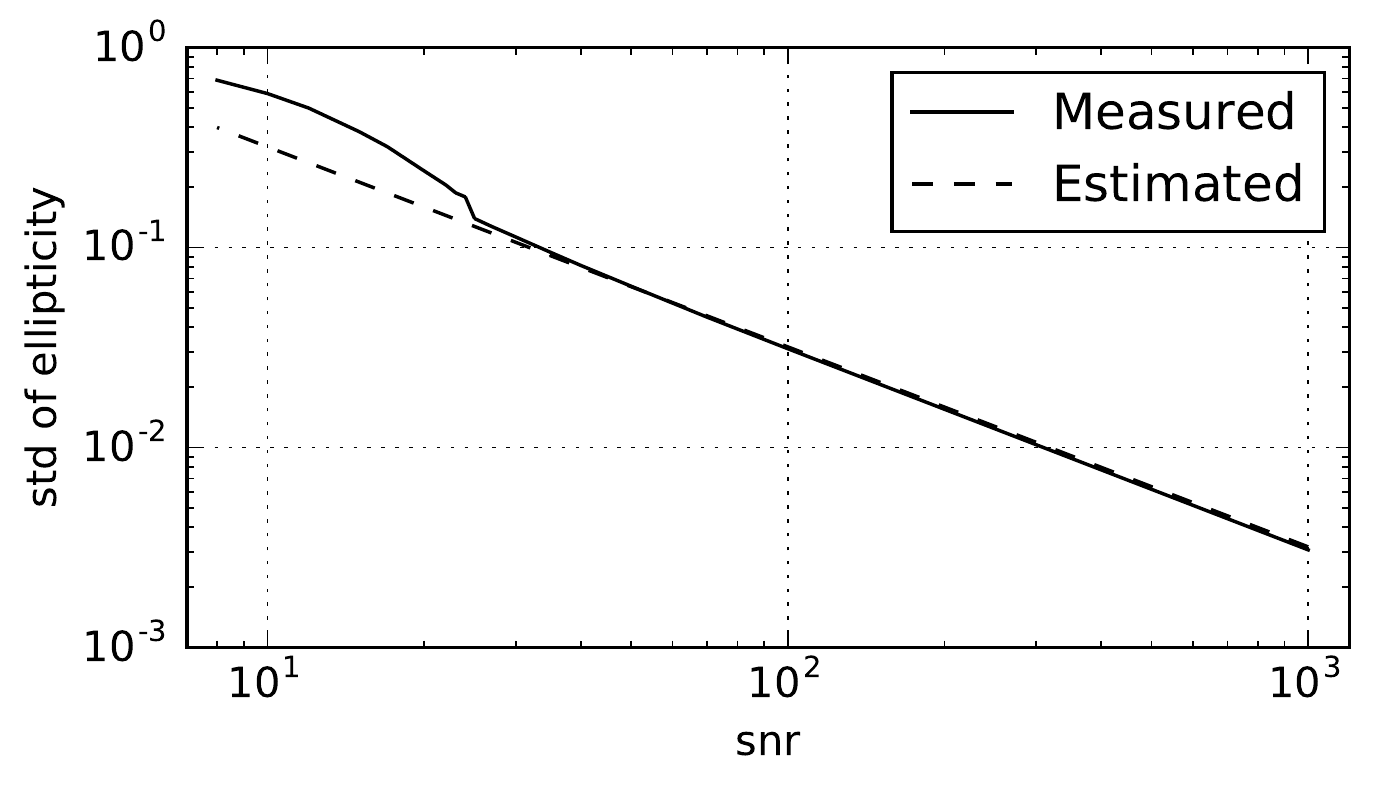}}
\caption{
\label{fig:sed_de}
Measured and estimated standard deviation of ellipticity as a function of SNR.
The horizontal and the vertical axis are SNR of galaxy and standard deviation of ellipticity, respectively.
The solid line and bashed line mean the measured and the predicted standard deviation, respectively. 
This is one of results of the simulations and in this figure  galaxy profile is Gaussian, PSF radius = 2.0 pixel and PSF ellipticity = 0.0.
}
\end{figure*}

Figure \ref{fig:sed_de} shows the measured and the estimated standard deviation of the ellipticity as a function of SNR.
We can see that the standard deviation can be predicted quite well especially in high SNR region.
We performed other simulations using several different parameters, and the results are similar 
with the one presented in this Figure .

\begin{figure*}[tbp]
\centering
\resizebox{1.0\hsize}{!}{\includegraphics{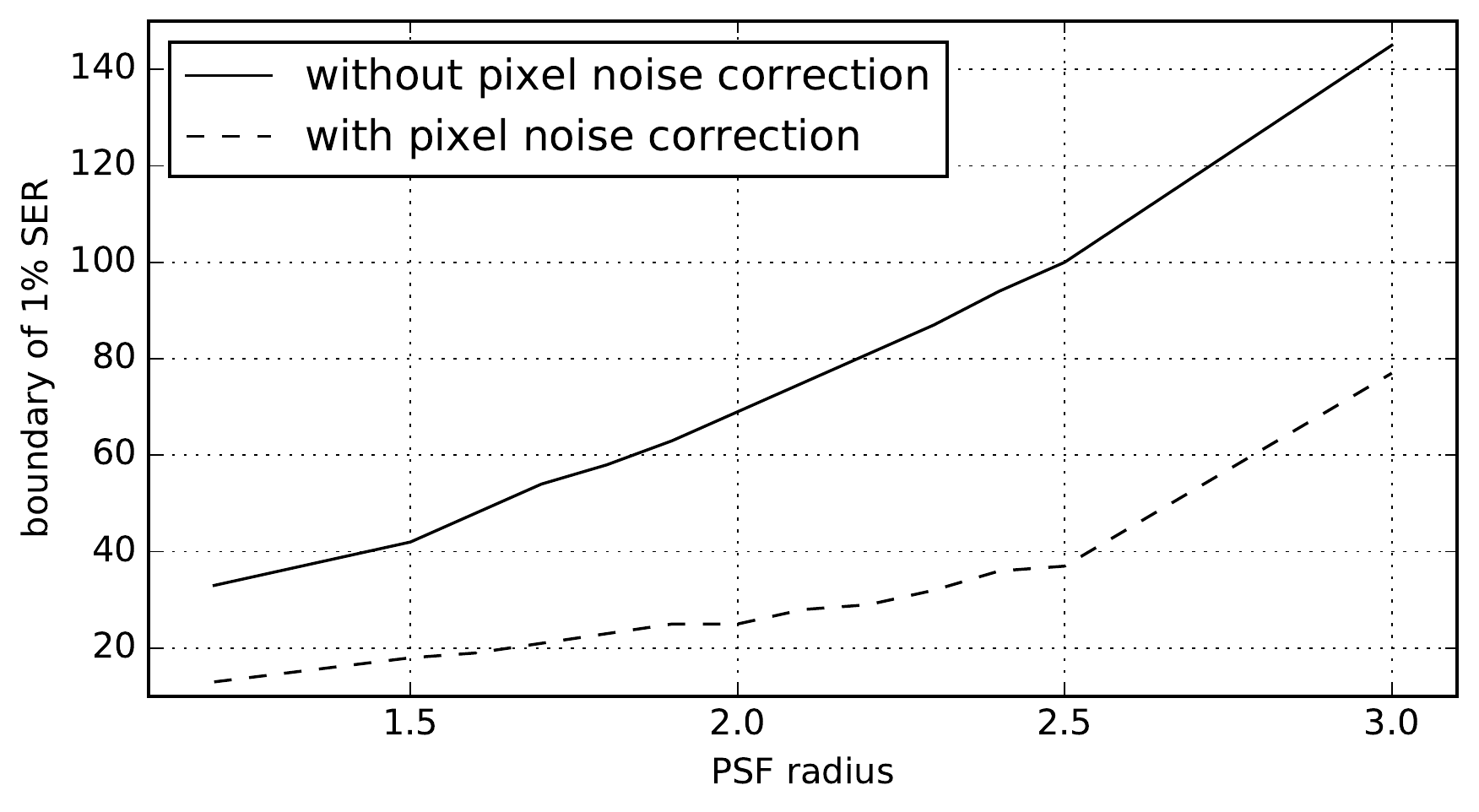}}
\caption{
\label{fig:SNRborderG}
SNR border of $1\%$ systematic error with and without the pixel noise correction for different PSF sizes.
The horizontal and vertical axis are PSF radius and SNR of the boundary, respectively.
The parameters used for this plot are that galaxy profile is Gaussian and PSF radius is 2.0 pixel, and 
PSF ellipticity = 0.0.
}
\end{figure*}

\begin{figure*}[tbp]
\centering
\resizebox{1.0\hsize}{!}{\includegraphics{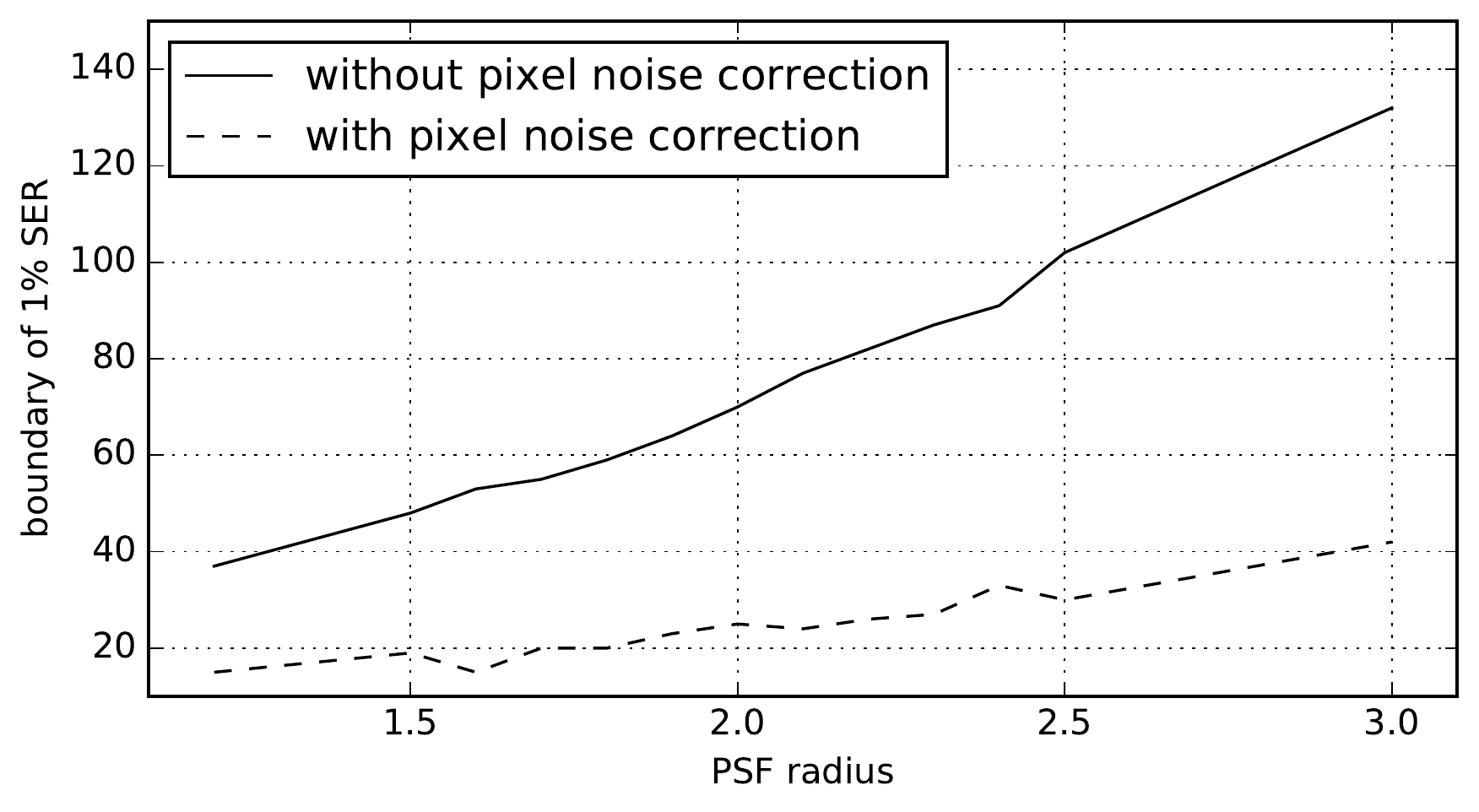}}
\caption{
\label{fig:SNRborderS}
same figure as figure \ref{fig:SNRborderG} except galaxy profile is Sersic.
}
\end{figure*}

Figure \ref{fig:mean_de_G_20} to \ref{fig:mean_de_S_01} show the systematic error ratio of  PSF corrected ellipticity and PSF and Pixel Noise corrected ellitpticity from true ellipticity as a function of SNR, where the vertical axis is the systematic error ratio $\rSE$ defined as $\rSE = |(\lrt{\mbe} - \mbetrue)/\mbetrue|$ where $\lrt{\mbe}$ is the mean value of ellipticity with or without pixel noise correction and $\mbetrue$ is the true ellipticity.
The results show that the pixel noise makes overestimation in shear analysis and it reaches $1\%$ around $\rSN = $[50,70,100] for PSF radius = [1.5, 2.0, 2.5] pixels and galaxy radius = 2.0 pixels,
but the Pixel Noise correction can correct $80 \sim 90\%$ of systematic error from the pixel noise especially in high SNR region, the  ``high'' region here means the region where the systematic error obeys inverse square low with respect to SNR.
Figure \ref{fig:SNRborderG} and \ref{fig:SNRborderS} show boundaries of $1\%$ systematic error ratio in SNR with and without pixel noise correction.

\begin{figure*}[tbp]
\centering
\resizebox{1.0\hsize}{!}{\includegraphics{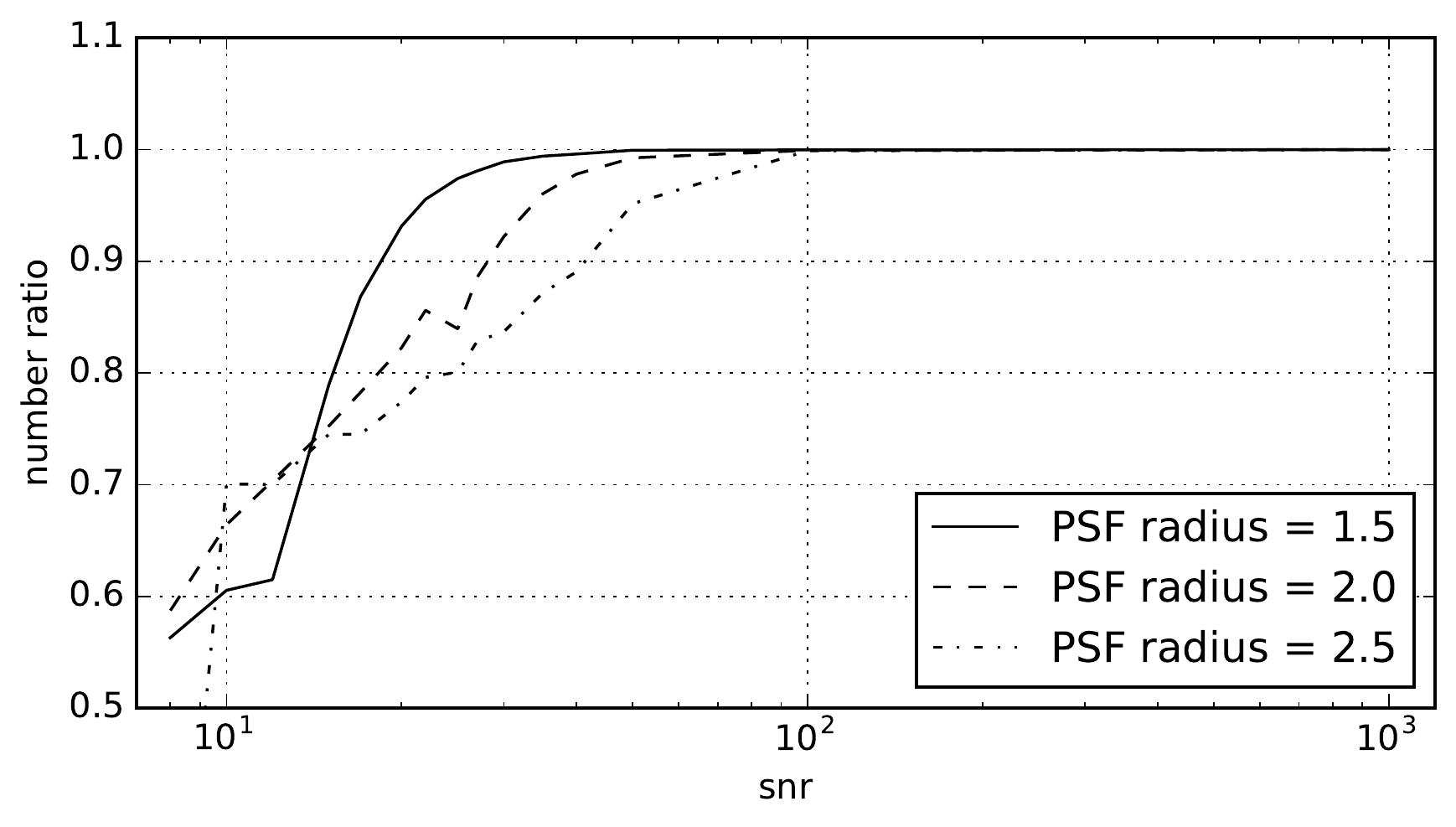}}
\caption{
\label{fig:reducednumberG}
The number ratio of galaxy which can be used for measuring shear as a function of SNR.
1.0 means all of galaxies are used for calculating the systematic error ratio.
PSF radius are 1.5(solid line), 2.0(dashed line) and 2.5(dot and dashed line) pixels.
}
\end{figure*}

\begin{figure*}[tbp]
\centering
\resizebox{1.0\hsize}{!}{\includegraphics{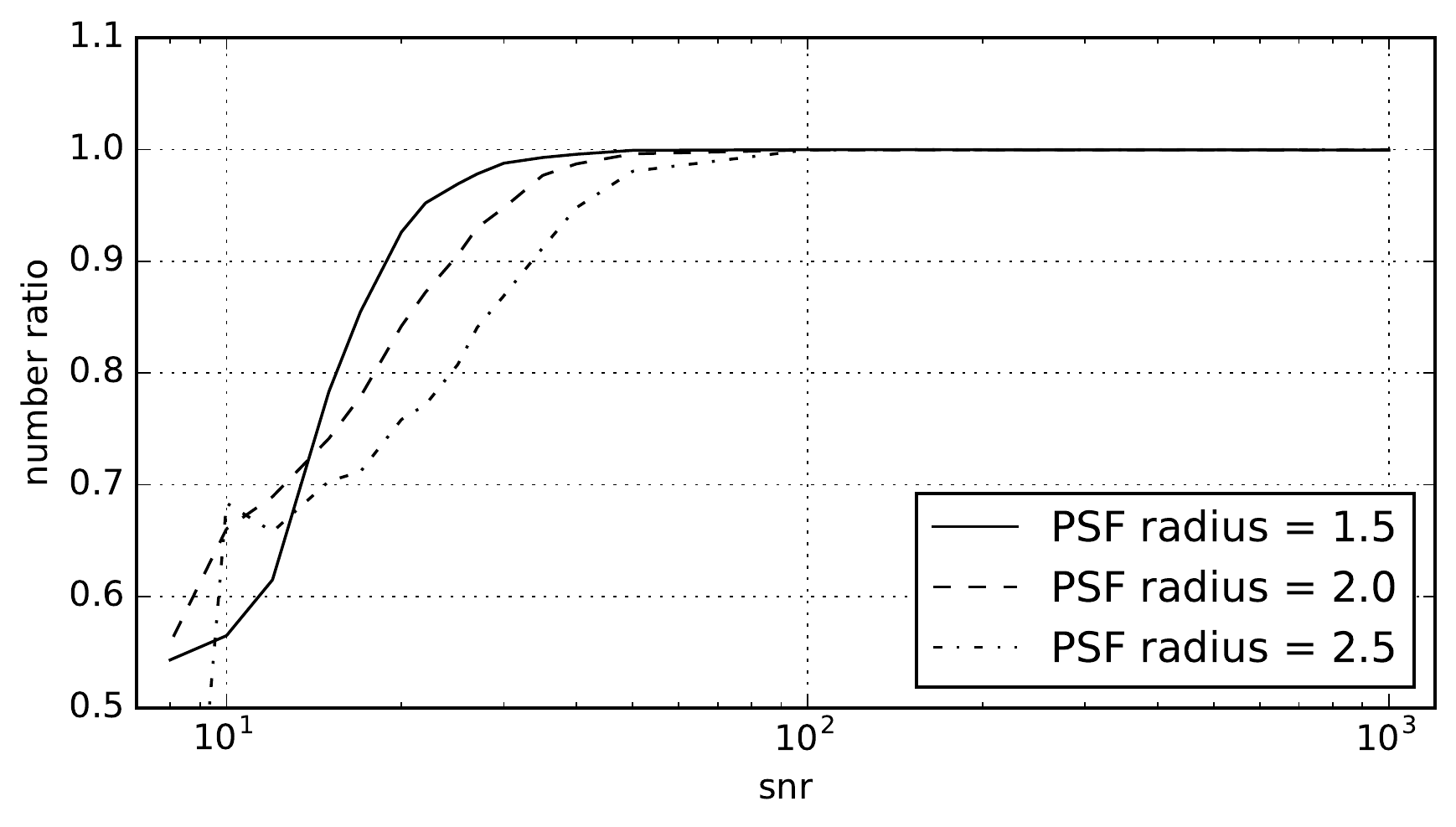}}
\caption{
\label{fig:reducednumberS}
Same figure as \ref{fig:reducednumberG} except galaxy profile is Sersic.
}
\end{figure*}

Figure \ref{fig:reducednumberG} and \ref{fig:reducednumberS} show the number ratio of galaxies we may  able to use for measuring the systematic error ratio as a function of SNR. 
In the low SNR region, some of images are rejected by some reasons such as the determined radius is too small and so on, and in this simulation we used 5$\sigma$ clipping for rejecting outlier due to the divergence of the correction value, so the lower SNR galaxies have higher rejected number.
The rejection number depends on the analysis parameters, so this figure shows just a sample in this simulation, but it is useful to know the typical number ratio of galaxy we can use in real analysis.
We can see $80\% \sim 90\%$ of galaxies can be used at $\rSN = 20$.
The divergence in the correction comes from the divergence of $\CoefP$, so it may be modified by 
adding certain value in $\CoefP$ to avoiding the divergence.
We will study the modification in detail in future works.

\begin{figure*}[tbp]
\centering
\resizebox{1.0\hsize}{!}{\includegraphics{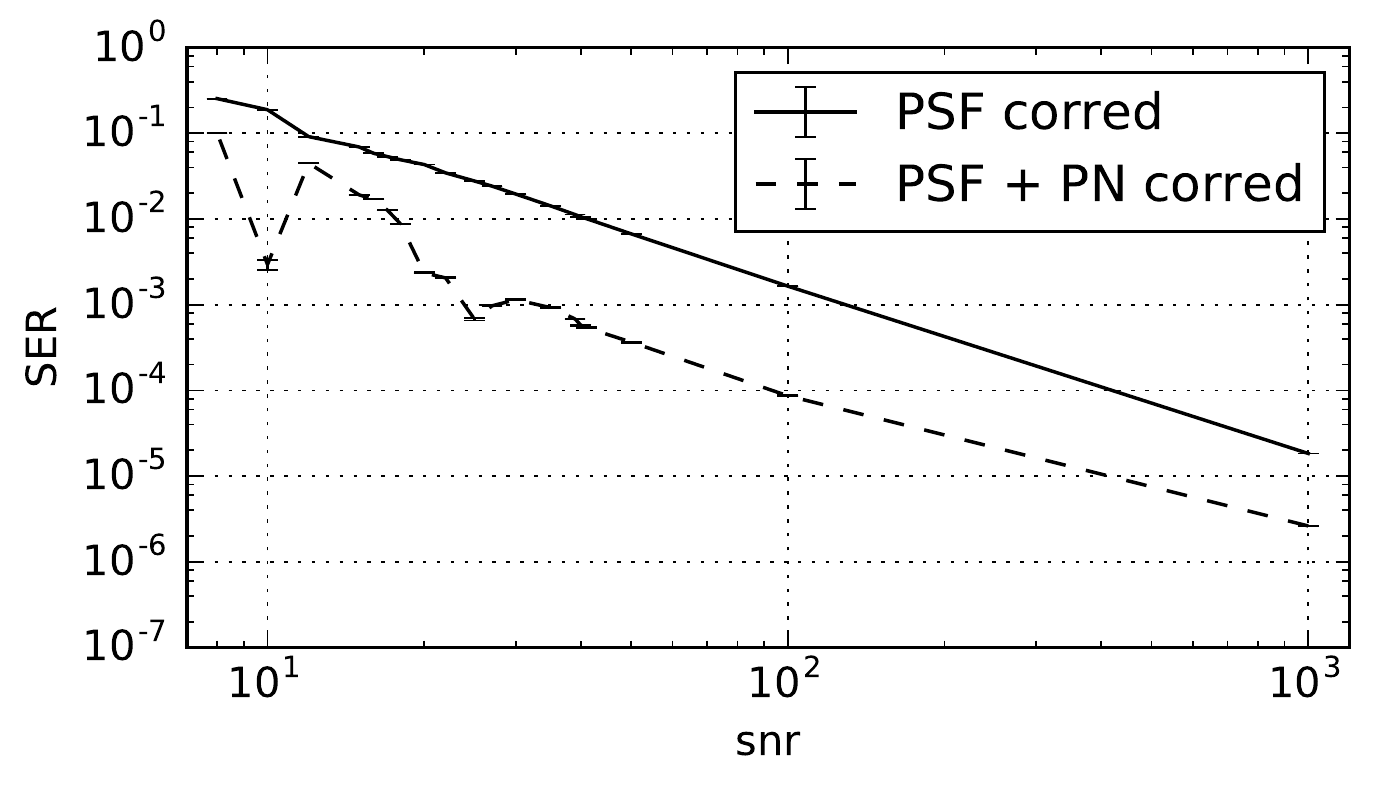}}
\caption{
\label{fig:mean_de_G_15}
The systematic error ratio of ellipticity without(solid line) and with(dashed line) the pixel noise correction.
In this simulation galaxy profile is Gaussian, PSF radius is 1.5 pixels and PSF ellipticity is 0.0.
The horizontal and the vertical axis are SNR of galaxy and systematic error ratio, respectively.
}
\end{figure*}
\begin{figure*}[tbp]
\centering
\resizebox{1.0\hsize}{!}{\includegraphics{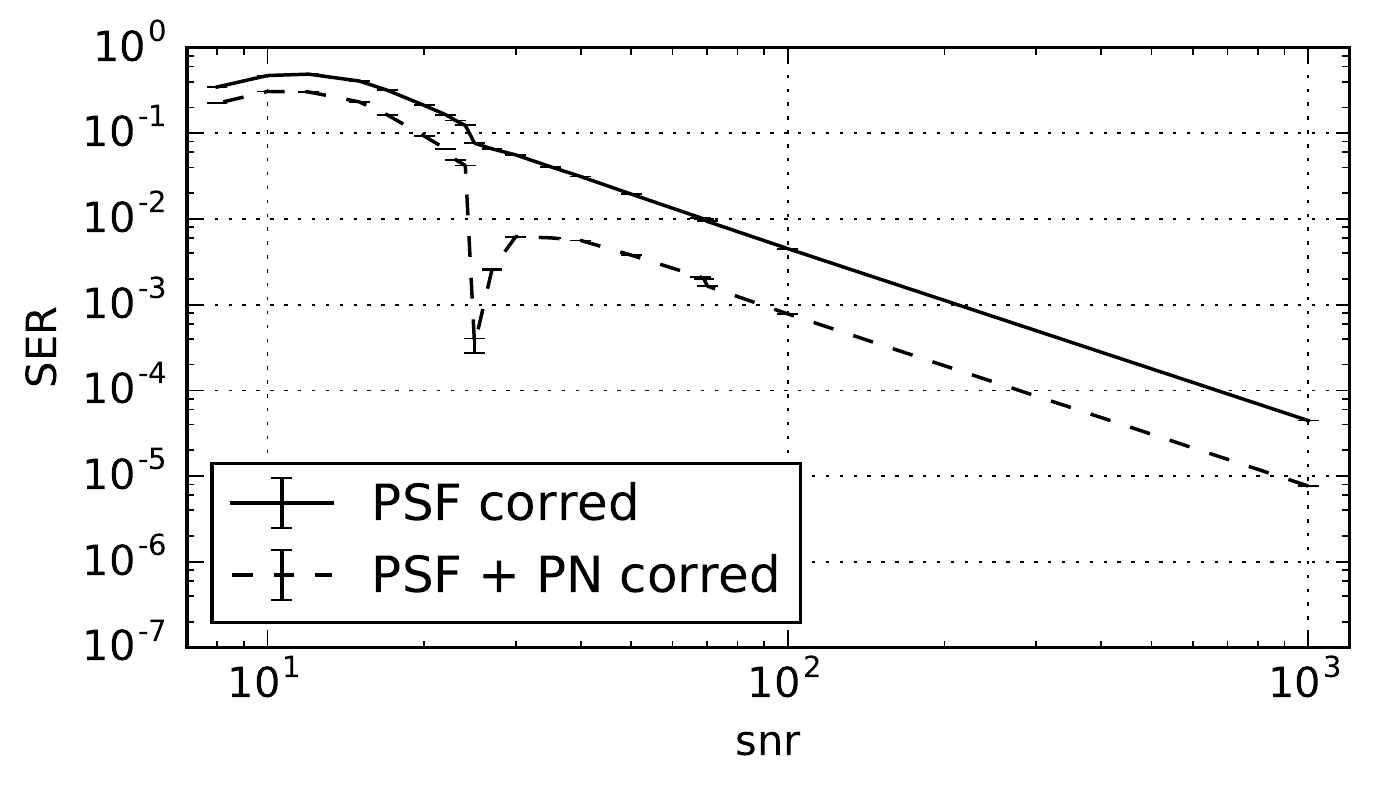}}
\caption{
\label{fig:mean_de_G_20}
Same figure as figure \ref{fig:mean_de_G_15}, except PSF radius = 2.0 pixel.
}
\end{figure*}
\begin{figure*}[tbp]
\centering
\resizebox{1.0\hsize}{!}{\includegraphics{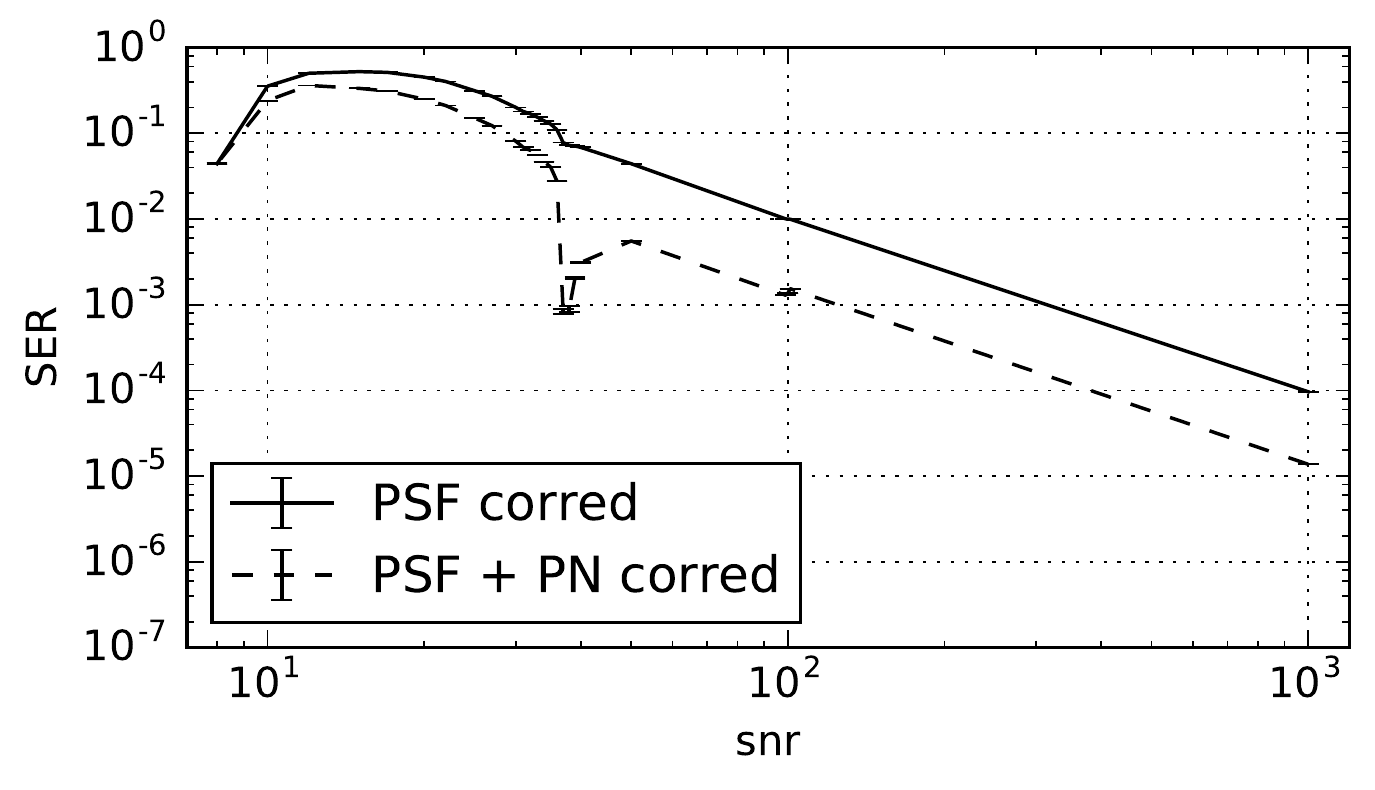}}
\caption{
\label{fig:mean_de_G_25}
Same figure as figure \ref{fig:mean_de_G_15}, except PSF radius = 2.5 pixel.
}
\end{figure*}
\begin{figure*}[tbp]
\centering
\resizebox{1.0\hsize}{!}{\includegraphics{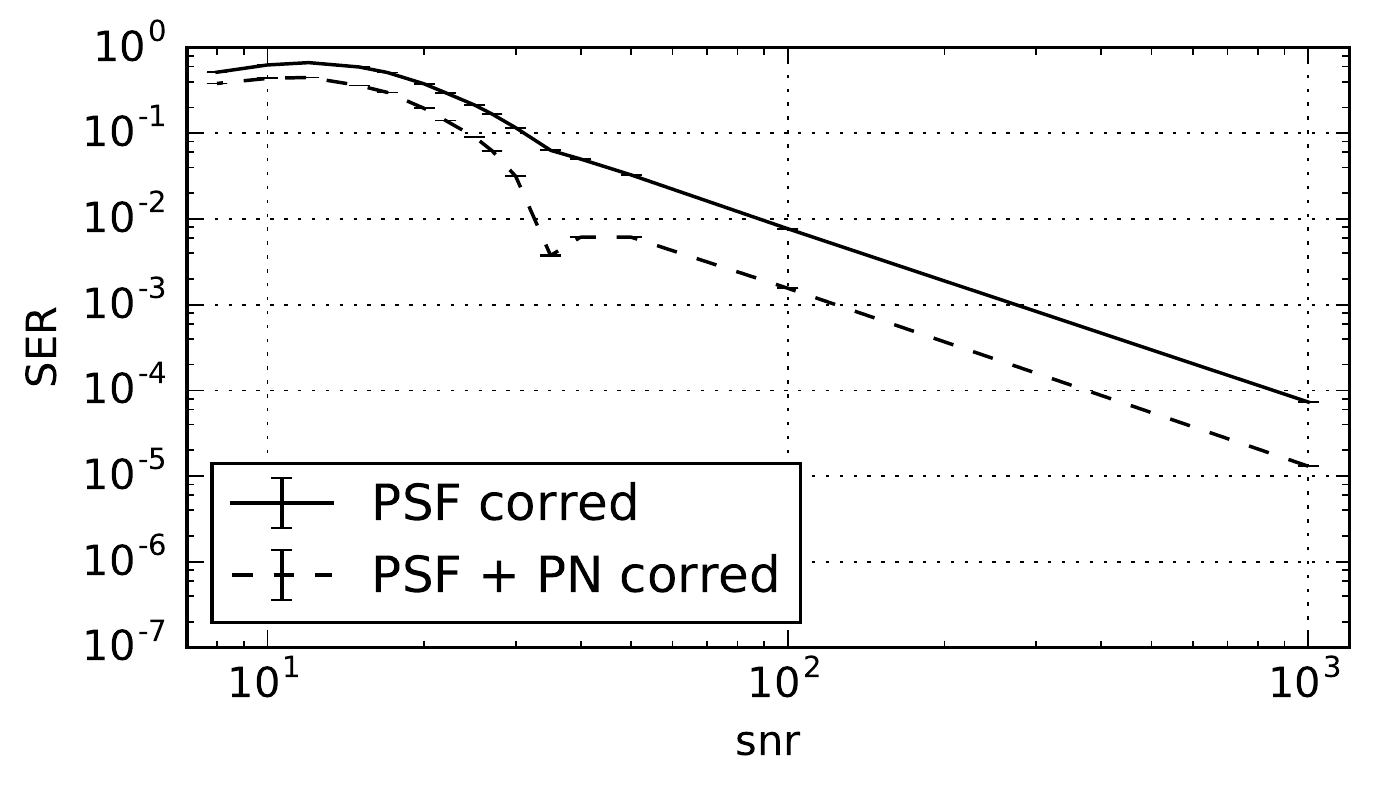}}
\caption{
\label{fig:mean_de_G_10}
Same figure as figure \ref{fig:mean_de_G_20}, except PSF ellipticity = -0.1.
}
\end{figure*}
\begin{figure*}[tbp]
\centering
\resizebox{1.0\hsize}{!}{\includegraphics{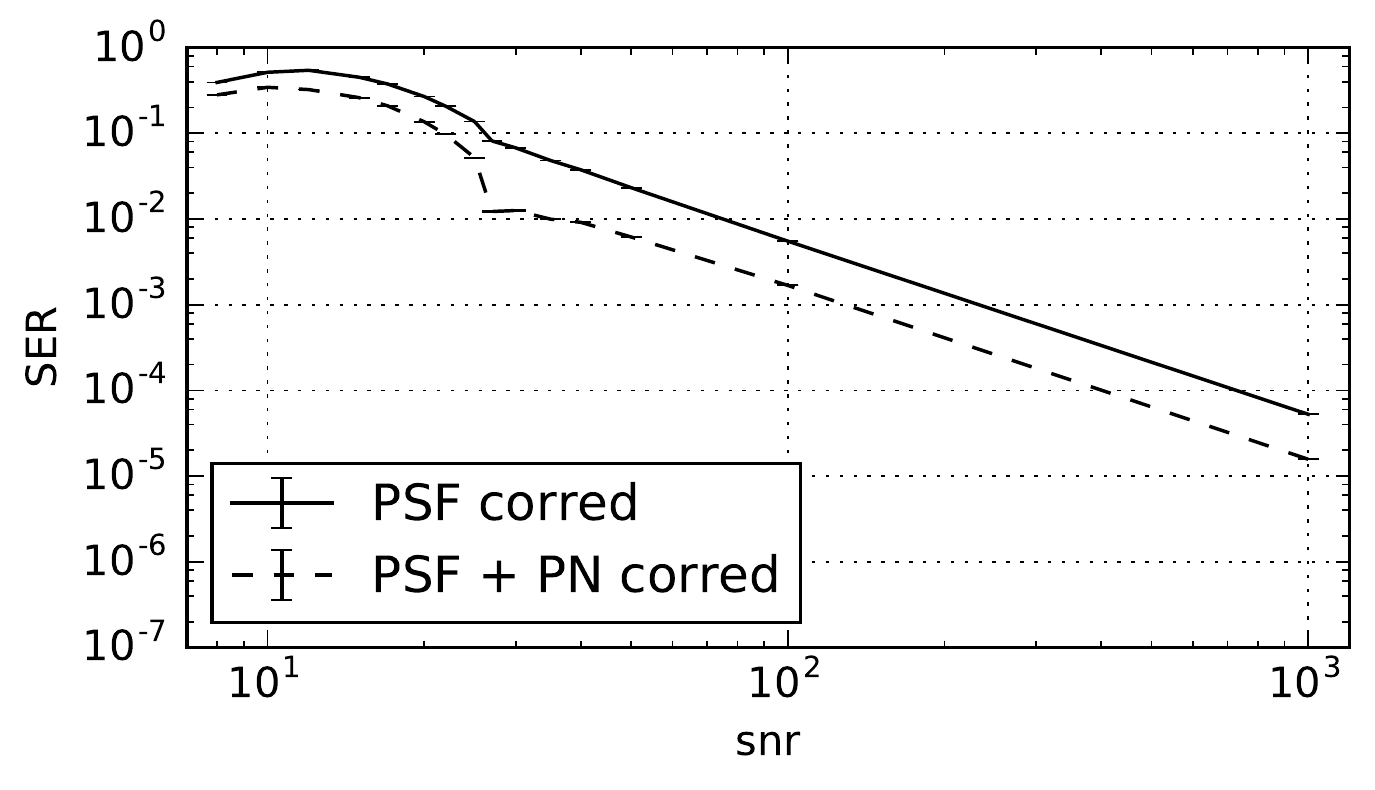}}
\caption{
\label{fig:mean_de_G_01}
Same figure as figure \ref{fig:mean_de_G_20}, except PSF ellipticity = 0.1i.
}
\end{figure*}

\begin{figure*}[tbp]
\centering
\resizebox{1.0\hsize}{!}{\includegraphics{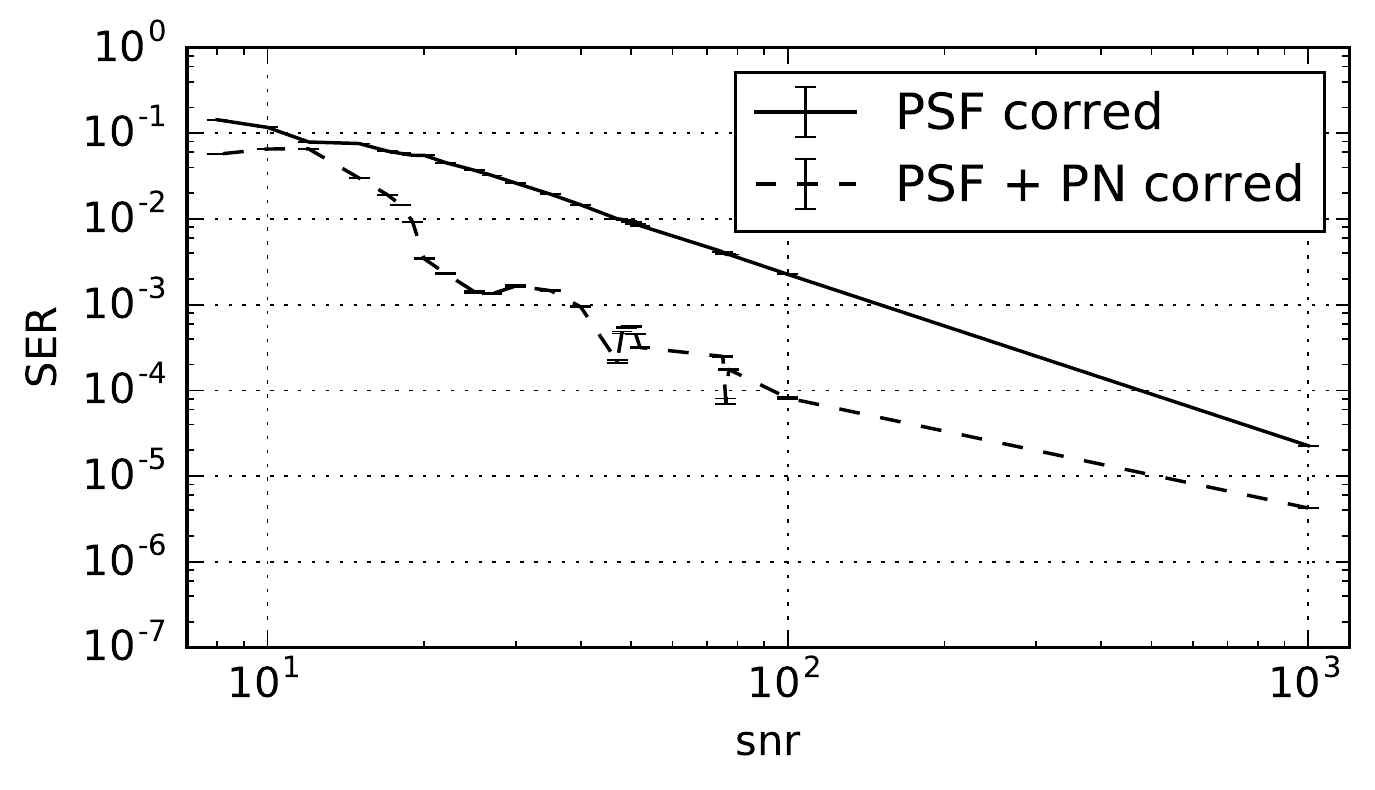}}
\caption{
\label{fig:mean_de_S_15}
Same figure as figure \ref{fig:mean_de_G_15}, except galaxy profile is Sersic.
}
\end{figure*}
\begin{figure*}[tbp]
\centering
\resizebox{1.0\hsize}{!}{\includegraphics{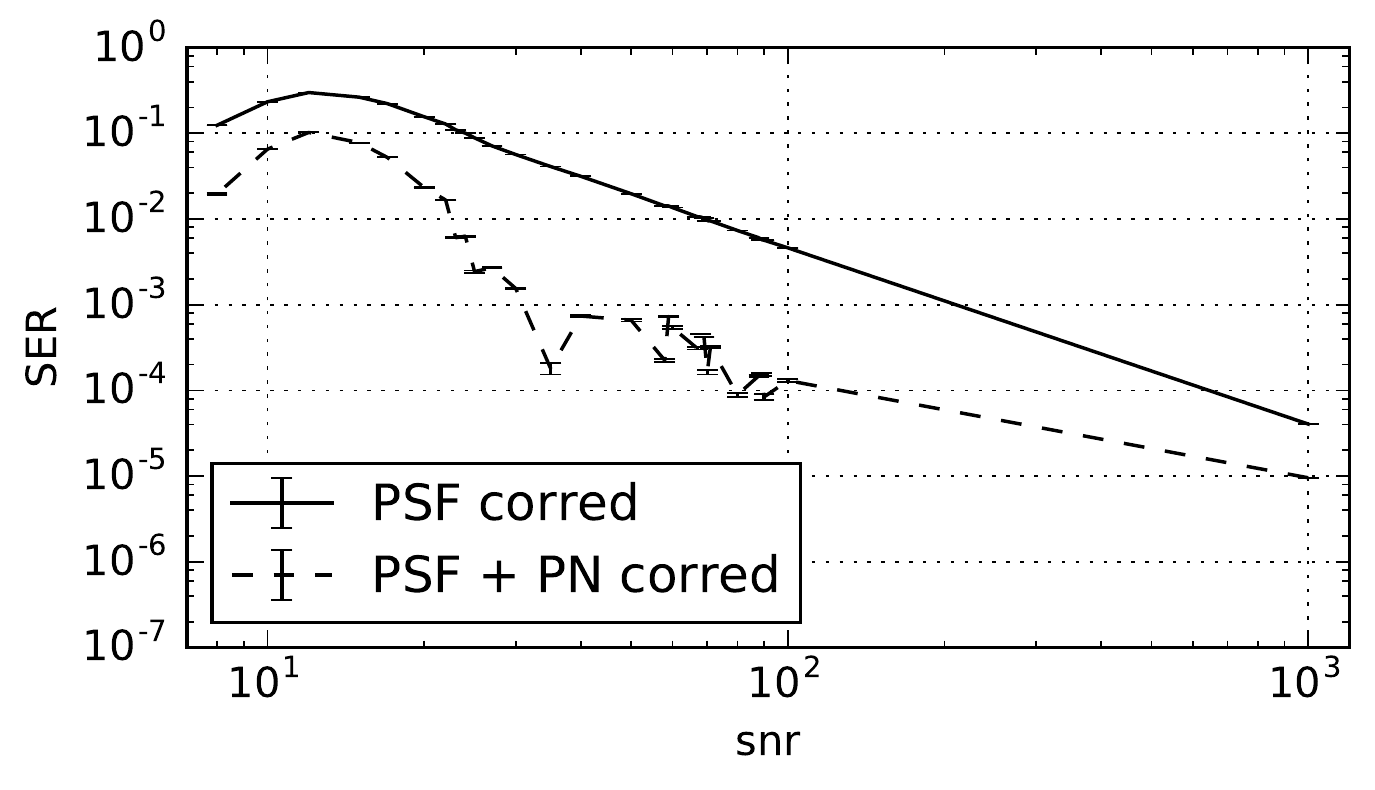}}
\caption{
\label{fig:mean_de_S_20}
Same figure as figure \ref{fig:mean_de_S_15}, except PSF radius = 2.0 pixel.
}
\end{figure*}
\begin{figure*}[tbp]
\centering
\resizebox{1.0\hsize}{!}{\includegraphics{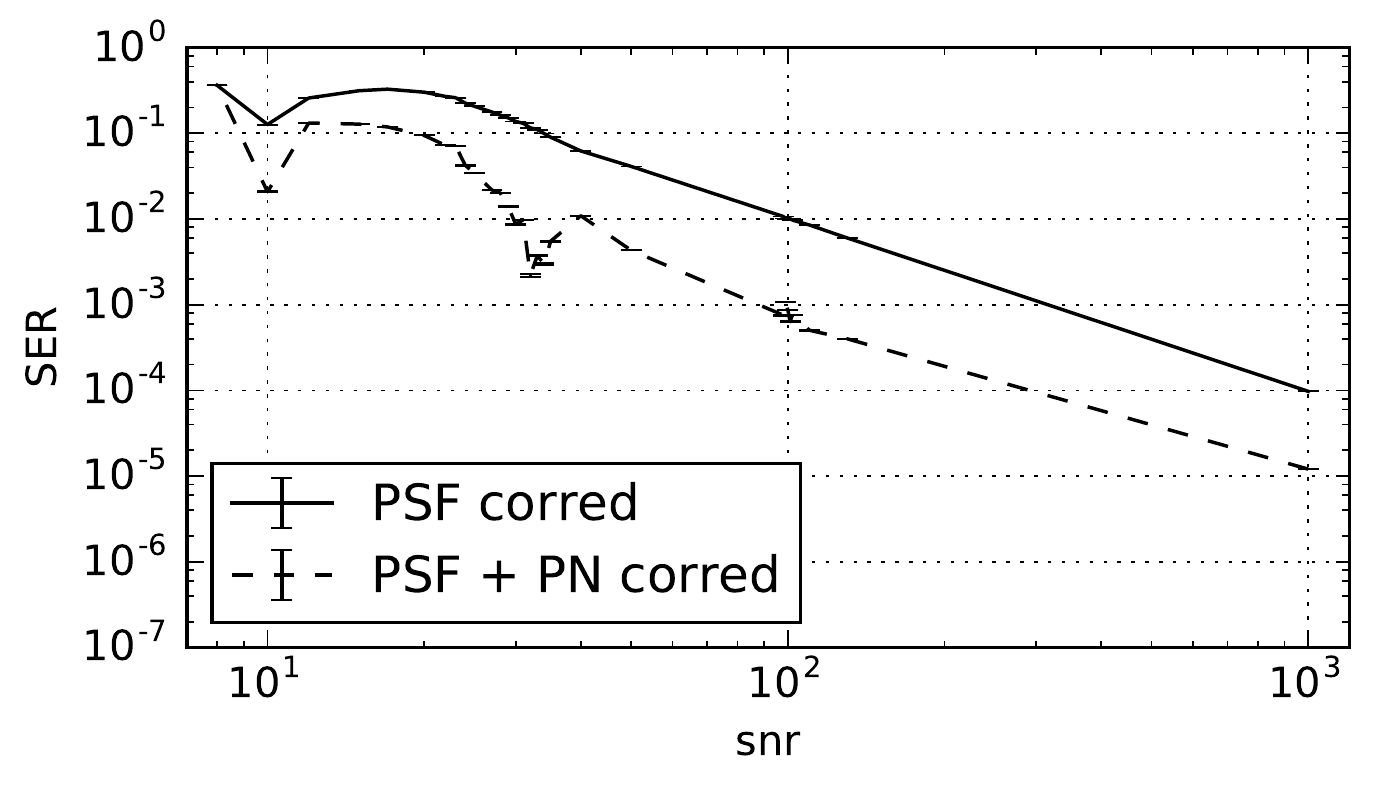}}
\caption{
\label{fig:mean_de_S_25}
Same figure as figure \ref{fig:mean_de_S_15}, except PSF radius = 2.5 pixel.
}
\end{figure*}
\begin{figure*}[tbp]
\centering
\resizebox{1.0\hsize}{!}{\includegraphics{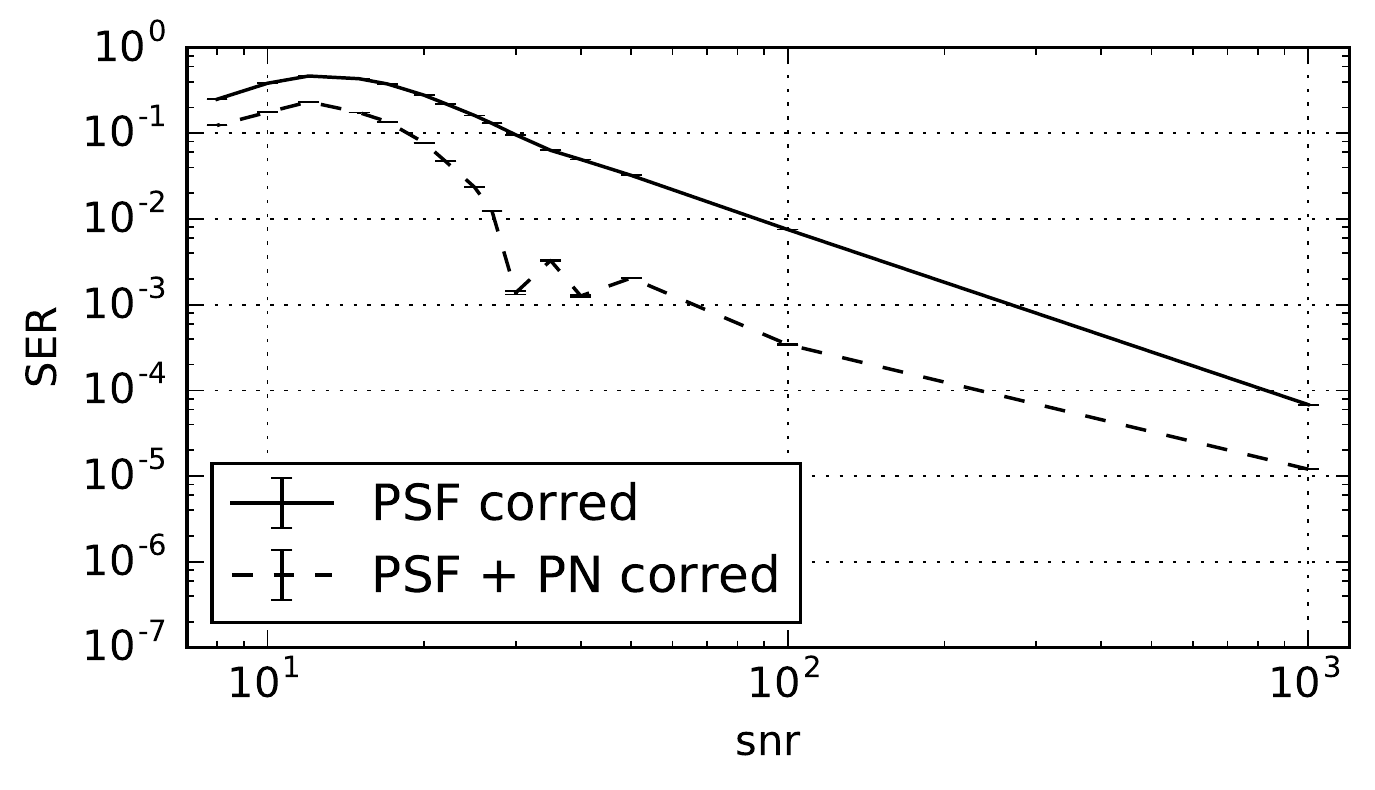}}
\caption{
\label{fig:mean_de_S_10}
Same figure as figure \ref{fig:mean_de_S_20}, except PSF ellipticity = -0.1.
}
\end{figure*}
\begin{figure*}[tbp]
\centering
\resizebox{1.0\hsize}{!}{\includegraphics{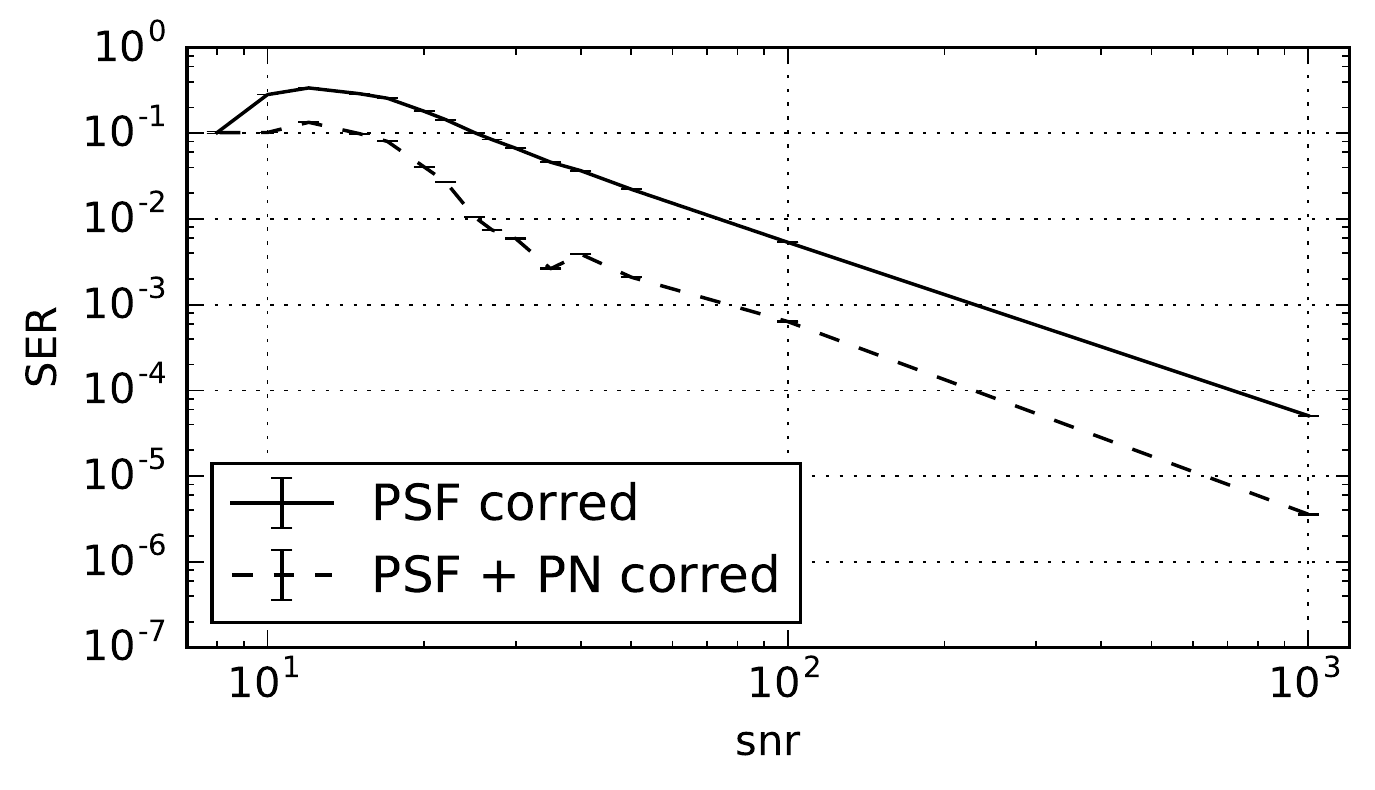}}
\caption{
\label{fig:mean_de_S_01}
Same figure as figure \ref{fig:mean_de_S_20}, except PSF ellipticity = 0.1i.
}
\end{figure*}



\section{Summary}

In this paper we have developed a new formulation of PSF correction in ``ERA'' to correct pixel noise effect. 
The basic idea is very simple, adding pixel noise in the equation of measuring PSF corrected ellipticity, then expanding the pixel noise up to the  2nd order in the deviation from the true ellipticity with assumption that the amplitude of the pixel noise count is small.
The 1st order pixel noise effect is random, so the mean value is 0 but it can be used to estimate standard deviation of ellipticity from pixel noise. 
The intrinsic ellipticity can be measured by the 1st pixel noise effect with the assumption that the  observed standard deviation has two components, and it is important to determine the weight for each galaxies.
The mean of the 2nd order pixel noise gives the systematic error and is very important for precise weak lensing shear analysis.

From the result of the simulation test, we can see that our correction method can correct $80\%\sim90\%$ of the systematic error by pixel noise within the parameter region we choose, and so we can use lower SNR galaxies such as $SNR\sim 20$ keeping the systematic error lower than $1\%$. This means that the method enables us to use more faint, i.e. more higher red-shift, galaxies with higher number density of background galaxy for weak gravitational lensing shear analysis. 
 
In the following paper we apply our method to HSC SSP wide field survey data 
 to estimate how much number density of galaxy in total and in each redshift bins can be increased by this correction method.

\acknowledgements
We would thank Erin Shelden very much for many useful discussions. 
This work is supported in part by a Grant-in-Aid for Science Research from
JSPS(No.17K05453 to T.F).


\end{document}